\documentclass{jfm}
\usepackage{graphicx}
\usepackage{psfrag, mathrsfs, subfig, amsmath, amssymb, natbib, multirow, bbding}

\DeclareFontFamily{OT1}{pzc}{}
\DeclareFontShape{OT1}{pzc}{m}{it}{<-> s * [1.10] pzcmi7t}{}
\DeclareMathAlphabet{\mathpzc}{OT1}{pzc}{m}{it}

\newcommand\pd[2]{\frac{\partial#1}{\partial#2}}

\newcommand\dv[2]{\frac{\ud #1}{\ud #2}}

\newcommand\ud{\textrm{d}}

\makeatletter
\newcommand\Rmnum[1]{\expandafter\@slowromancap\romannumeral #1@}
\makeatother

\title[Dynamical dispersion in turbulent plumes]
  {Dynamical dispersion in turbulent plumes}

\author[M.~M.~Scase \& A.~W.~Woods]
{Matthew M.~Scase$^{1, 2}$  \& Andrew W.~Woods$^{1, 3}$}

\affiliation{$^1$Geophysical Fluid Dynamics, Woods Hole Oceanographic Institution, Woods Hole, MA 02543, USA\\
$^2$School of Mathematical Sciences, University of Nottingham, Nottingham NG7 2RD, UK\\
$^3$BP Institute, University of Cambridge, Madingley Road, Cambridge CB3 0EZ, UK}

\date{\today} 

\begin{document}

\maketitle

\begin{abstract}
Experimental observations of turbulent buoyant plumes, produced by a constant source of buoyancy, have been described with great success using a horizontally averaged model for the conservation of mass, momentum and buoyancy flux.  However, experimental observations of plumes with time-dependent buoyancy fluxes has proved more challenging for quantitative models. At each level in the plume, the horizontal variation in velocity leads to an along-axis shear and hence dispersive transport relative to the mean.  With a time-dependent source, axial dispersion of the dynamic properties of the plume has a key role in the evolution of the flow. Using ideas of mixing length theory, we introduce a model for this axial dispersion, and test the model by comparison with experimental observations of plumes in which the buoyancy flux is suddenly decreased or suddenly increased. In both cases, we show the transition from one buoyancy flux to another is self-similar, and using the data, we find the axial dispersion may be expressed as $\beta U b$ where $\beta$ lies in the range $0.70$--$0.88$ and $U$ and $b$ are respectively the horizontally averaged vertical velocity and radius of the plume at height $z$ and time $t$.  Our model also reduces to that of a classical plume when the buoyancy flux is steady
\end{abstract}


\section{\label{sec:int}Introduction}


Turbulent buoyant plumes arise in many environmental and geophysical situations where there is a localised source of buoyancy. In the special case in which there is  a steady source of buoyancy,   Morton Taylor \& Turner (1956) tested the predictions of an integral model for the conservation of mass, momentum and buoyancy in a plume, coupled with an entrainment hypothesis, by using the results of a  series of laboratory experiments. Applications ranging from volcanic eruption columns \citep[see e.g.,][]{woods88, scase09}, ocean currents (see e.g., Holland 2011; Holland, Hewitt \& Scase 2014), hydrothermal plumes, fire plumes through to convective plumes in buildings have been studied, advancing this original work \citep{woods10}.

In some situations, the buoyancy flux may vary in time, and this leads to interesting questions about how the evolving buoyancy flux migrates through the plume and provides a new test for the model, and in particular the entrainment hypothesis. In this context, \citet{scase06}, Scase, Caulfield \& Dalziel (2008), Scase, Aspden \& Caulfield (2009) and \citet{scaseHewitt12} have explored the dynamics of plumes in which the source buoyancy flux changes suddenly from one constant value to another. \citet{scd08} present a series of laboratory experiments which identify that the change in source conditions leads to a region of adjustment which migrates through the plume, such that near the source a steady plume with the new buoyancy flux becomes established, while further from the source, the flow adjusts over some region to that of the original plume.  However, the theoretical description of this transition region has to date posed some challenges (e.g., Scase \& Hewitt 2012; Woodhouse, Phillips \& Hogg 2016). 

Parallel to this work, there has been some investigation of the mixing of passive tracer within plumes and jets. Landel, Caulfield \& Woods (2012) and \citet{roccoWoods15} demonstrated that in a steady jet or plume, a pulse of tracer becomes axially dispersed owing to the variation in the velocity with horizontal position across the flow. Guided by the ideas of shear dispersion, they demonstrated that this axial dispersion could be described by using an axial dispersivity based on mixing length theory. In that work, the tracer was dynamically passive. In the present contribution, we explore the effect of such dispersion on the dynamically active properties of the flow, and test the model using the experimental data presented by \citet{scd08, sac09}.

In \S\,\ref{sec:model}, we propose a model of a time dependent plume, including an idealised model of the axial dispersion. We demonstrate that the model is analogous to the classical model of turbulent buoyant plumes when there is a steady buoyancy flux, and we then illustrate that the model admits similarity solutions to describe the transition in the flow following a change in the source buoyancy flux from one value to a second. We re-analyse the experimental data originally presented by \citet{scd08, sac09} and demonstrate that the transition region does in fact evolve in a self-similar manner.  By comparison with our model, we determine the empirical constant $\beta$, which constrains the magnitude of the axial dispersivity, $\beta U b$, where $U$ and $b$ are the characteristic vertical plume velocity and plume radius. 
Finally, we discuss some possible future avenues for research.


\section{\label{sec:model}Model}


\begin{figure}
\begin{center}
\includegraphics{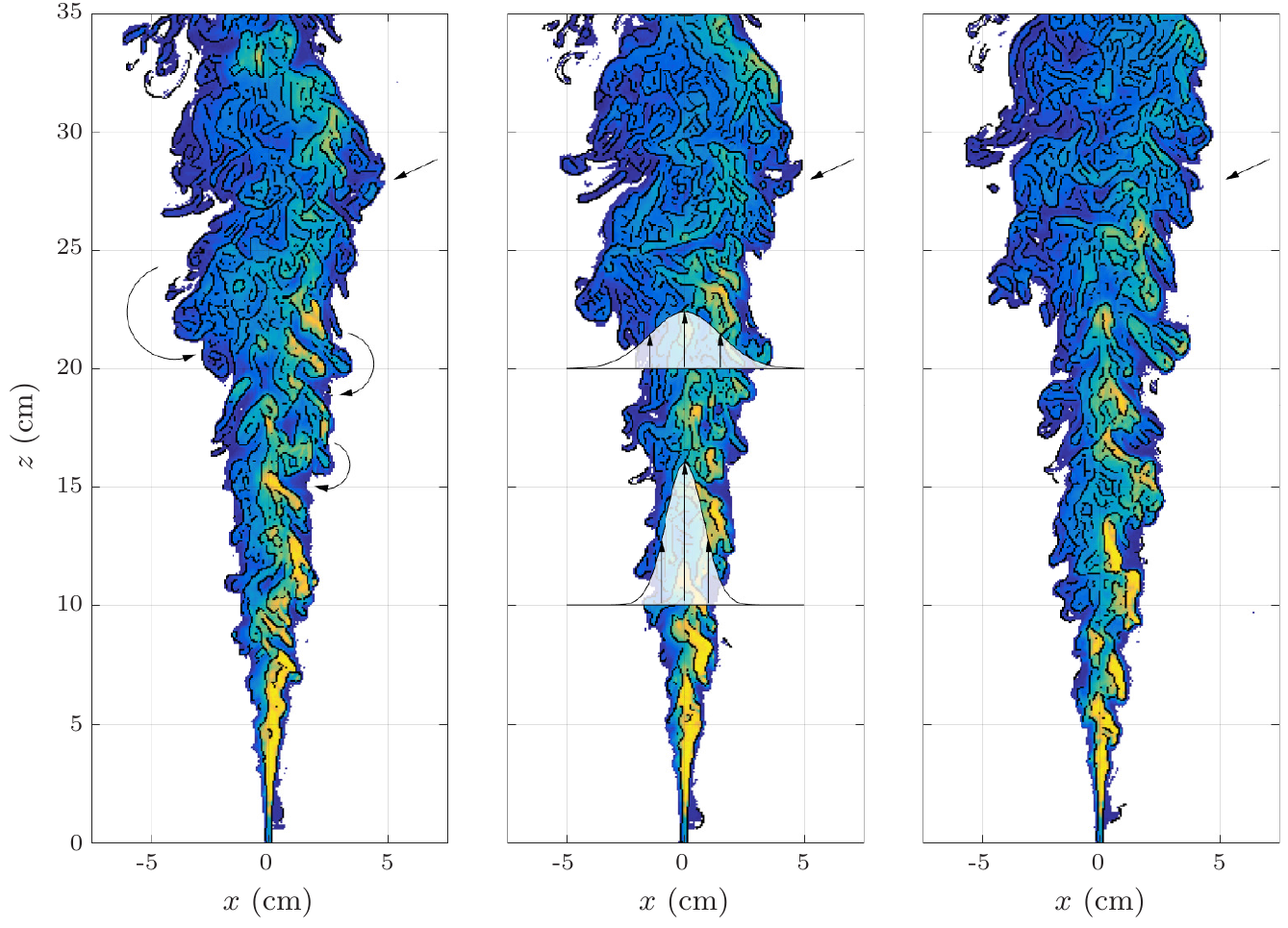}
\end{center}
\caption{\label{fig:cartoon}Image of a plume rising through a homogeneous ambient fluid.  The colours represent buoyancy.  Fluid at the centre of the plume rises more rapidly than fluid at the edges of the plume leading to a Taylor-like shear dispersion.  The images shown are from Movie 2 and correspond to times $t=-8.417$\,s, $t = -8.083$\,s and $t=-7.750$\,s.}
\end{figure}

For a Boussinesq flow, the governing equations for the rate of change in time of the horizontally integrated mass, momentum and buoyancy  may be expressed in terms of the gradient of the horizontally integrated mass, momentum and buoyancy fluxes, combined with (i) a model for the entrainment of ambient fluid, (ii) the buoyancy force and (iii) a model to represent the axial dispersion associated with the turbulent mixing coupled with the shear \citep[cf.][]{LandelEtAl}. In building the model, we work with ensemble averages, so that the properties are averaged over the turbulent fluctuations. For each of the properties $f$ of the flow, we write $f = \bar f + \hat f$, where $\bar f$ is the ensemble average, and $\hat f$ denotes the turbulent fluctuations. 
The horizontally integrated, ensemble averaged  properties for the vertical velocity $u(r, \theta, z, t)$ and buoyancy $g'(r, \theta, z, t)$ are then given by 
\begin{equation}
\renewcommand{\theequation}{\arabic{section}.\arabic{equation}\,$a$,$b$}
\pi  U b^2  				= Q 		= \int \bar{u} \, \ud A, \qquad
\pi S_m U^2 b^2 			= S_m M	= \int \left(\bar{u}^2+\hat{u}^2\right) \ud A,
\end{equation}
\begin{equation}
\renewcommand{\theequation}{\arabic{section}.\arabic{equation}\,$c$}
\addtocounter{equation}{-1}
\pi S_f U G' b^2			 	= S_f F 	= \int \left(\bar{u} \bar{g}' + \hat{u}\hat{g'}\right) \ud A,
\end{equation}
where $U$, $b$ and $G'$ are the characteristic vertical velocity, radius and buoyancy of the plume, such that the ensemble-averaged volume, momentum and buoyancy, per unit height, are given by $\pi b^2$, $\pi b^2 U$ and $\pi b^2 G'$.  The shape factors $S_f$ and $S_m$ are associated with the radial distribution of the velocity and buoyancy \citep[cf.][]{turner79}.
\renewcommand{\theequation}{\arabic{section}.\arabic{equation}}

The dispersive flux arises from  the turbulent fluctuations in the flow coupled with the axial shear, and, using mixing length arguments for such a highly turbulent flow, we expect the associated dispersivity to be proportional to the characteristic speed and radius of the flow, $\beta U b$ \citep[cf.][]{scaseHewitt12, roccoWoods15}, where $\beta$ is a constant. This leads to fluxes of the horizontal integral of the dispersive buoyancy and momentum of the form $-\beta U b\ \partial (G'   b^2)/\partial z$ and  $-\beta U b\ \partial(U b^2)/\partial z$ respectively. As well as the dispersive transport of momentum and buoyancy, the eddies tend to transport material near the edge of the plume outwards and downwards relative to the mean flow, engulfing ambient fluid.  The mixture is then swept back into the plume, driven by the shear and becomes incorporated into the plume leading to the turbulent entrainment of ambient fluid into the plume.   Indeed, figure \ref{fig:cartoon} shows three instantaneous images of the concentration field in a turbulent plume taken in a plane which cuts through the centre of the plume and which is perpendicular to the line of sight.  The plume has a constant source of buoyancy.  The frames are taken from the experiment shown in Movie 1 at times $t=-8.417$\,s, $t = -8.083$\,s and $t=-7.750$\,s.  The plume's concentration field has been passed through an edge-detection algorithm \citep{canny86} to highlight the coherent structures within the flow and this is presented in Movie 2.  

 \citet{mtt56} proposed that the volume flux of  ambient fluid which is entrained into the plume is proportional to the vertical speed of the plume and the plume circumference, and this model is quantitatively consistent with experimental data.  However, the turbulent eddies also tend to smooth out discontinuities in the width of the plume, since eddies which form in the wider parts of the plume partially mix back into the adjacent narrower parts of the plume, owing to the shear in the flow.  This can lead to a net dispersive transport of volume in regions where there is a vertical gradient of radius, and, for consistency with the momentum and buoyancy transport, we hypothesize that this dispersive volume flux has the form $-\beta Ub\ \partial(b^2)/\partial z$. 

We now write down a model for the time-dependent plume, and then develop solutions of this model to test for consistency with (i) the classical steady plume solutions presented by \citet{mtt56} and (ii) the experimental data monitoring the progression of the transition region following  a discrete increase or decrease in the plume buoyancy flux \citep{scd08, sac09}. 
Combining the processes above, we hypothesize that the mass, momentum and buoyancy fluxes are governed by the relations 
\begin{subequations} \label{eq:sys}
\begin{equation} \label{eq:massCons}
\pd{}{t}\left(\frac{Q^2}{M}\right) + \pd{Q}{z} = 2\alpha M^{1/2} + \beta\pd{}{z}\left[M^{1/2}\pd{}{z}\left(\frac{Q^2}{M}\right)\right],
\end{equation}
\begin{equation} \label{eq:momCon}
\pd{Q}{t} + S_m\pd{M}{z} = \frac{QF}{M} + \beta\pd{}{z}\left(M^{1/2}\pd{Q}{z}\right),
\end{equation}
\begin{equation} \label{eq:buoyCon}
\pd{}{t}\left(\frac{QF}{M}\right) +S_f\pd{F}{z} = \beta\pd{}{z}\left[M^{1/2}\pd{}{z}\left(\frac{QF}{M}\right)\right].
\end{equation}
\end{subequations}


\subsection{Similarity solutions for a steady plume}

\begin{figure}
\begin{center}
\includegraphics{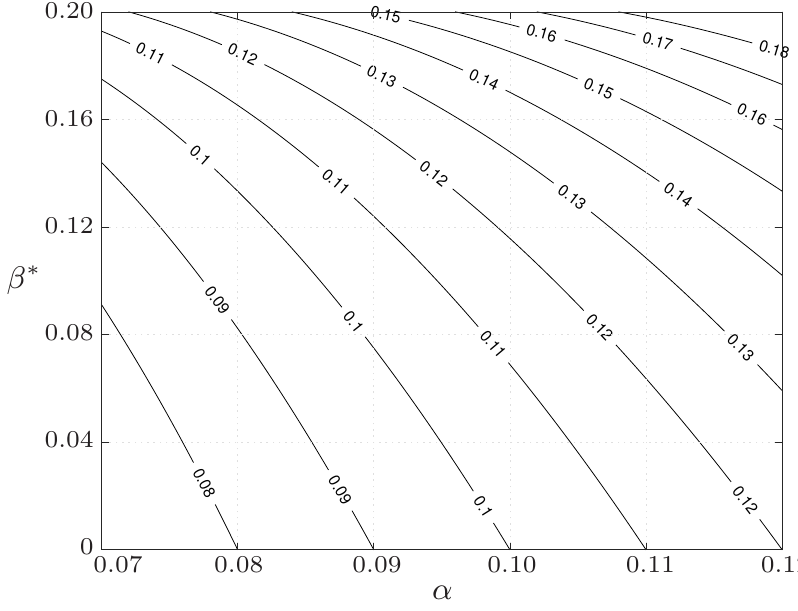}
\end{center}
\caption{\label{fig:abc}Contours of the classical entrainment coefficient $\alpha_c$ for varying entrainment coefficent $\alpha$ and scaled dispersion coefficient $\beta^* = 2\alpha\beta$.}
\end{figure}

A steady plume, generated by a constant source of buoyancy issuing from a point source follows a self-similar structure as a function of height \citep{mtt56}. The time-dependent system of equations \eqref{eq:sys} should admit steady solutions of the same form.  We therefore seek solutions
\begin{equation}
Q = (2\alpha)^{4/3}F_0^{1/3}z^{5/3} q_0, 
\quad
M = (2\alpha)^{2/3}F_0^{2/3}z^{4/3} m_0, 
\quad
F = F_0,
\end{equation}
where $F_0$ is the source buoyancy flux and $\alpha$ is the entrainment coefficient.  Consistency conditions following from \eqref{eq:sys} then require that 
\begin{equation} \label{eq:q0m0}
q_0 = \frac{12S_m m_0^2}{20\beta^* m_0^{3/2} + 9}
\quad\textnormal{and}\quad
m_0 = \left\{
\frac{9}{40}\frac{S_m\left[1-\left(1-24\beta^*/5\right)^{1/2}\right]-2\beta^*}{\beta^*\left[S_m(6S_m/5 - 1)+\beta^*\right]}\right\}^{2/3},
\end{equation}
where $\beta^* = 2\alpha\beta$.  The plume radius is defined as
\begin{equation} \label{eq:plumeRadius}
b = \frac{Q}{M^{1/2}} = 2\alpha z\frac{q_0}{m_0^{1/2}} = \frac{6\alpha z}{5}\frac{2}{1 + \left(1 - 24\beta^*/5\right)^{1/2}} ,
\end{equation}
this solution requires that $\beta^*<5/24$, and so we can write $b = 6\alpha z/5\left(1 + 5\beta^*/5 + O(\beta^{*2})\right)$.
This is analogous to the classical result 
\begin{equation}
b =  \frac{6\alpha_c z}{5}
\end{equation}
provided the classical entrainment coefficient $\alpha_c$ is related to the present entrainment coefficient, $\alpha$, and dispersion coefficient, $\beta^*$ according to
\begin{displaymath}
\alpha_c = \alpha \left[1 + \frac{6\beta^*}{5}+O\left(\beta^{*2}\right)\right]
\end{displaymath}
The equivalence of the two models implies that the present entrainment coefficient $\alpha$ is a fractionally smaller than the classical value \citep{mtt56} owing to the dispersion which in fact has the effect of carrying some of the entrained material to points in the plume ahead of the height at which it was entrained: the classical entrainment coefficient, which implicitly includes such dispersion, therefore appears a little larger. The magnitude of this difference can only be determined by measurement of the dispersion coefficient $\beta$ which forms the subject of the main part of this paper.  Figure \ref{fig:abc} shows contours of the equivalent classical entrainment coefficient $\alpha_c$ for varying entrainment coefficient $\alpha$ and scaled dispersion coefficient $\beta^* = 2\alpha\beta$.  
For a steady source of buoyancy, the dispersive flux of buoyancy exactly vanishes so that the buoyancy flux at each height is a constant. The dispersive transport of mass is a constant fraction of the entrained mass flux, while the dispersive transport of momentum scales with height exactly as the buoyancy force.  Note also that the plume radius \eqref{eq:plumeRadius} is independent of both shape factors $S_m$ and $S_f$.

The effects of the axial dispersion can be much more important on the evolution of flows in which the source is  time-dependent. As mentioned in the introduction, we test the new model \eqref{eq:sys} by examining solutions for the case in which the steady flux supplying the plume changes from one value to another. There is already detailed experimental data available for the this problem \citep{scd08, sac09}, and we  use the data to test and calibrate the present model.

\subsection{Similarity solutions for plumes involving a sudden change in buoyancy flux}


Guided by the classical solutions for a steady state plume, we now seek solutions to describe the evolution of the transition region in which a plume adjusts from one steady buoyancy flux to another. Far from the region of transition, we expect the flow to asymptote to the solution for a steady plume with constant buoyancy flux. These limiting solutions have the same structure but a different value of the buoyancy flux and so given that there is no external length scale regulating the along-plume extent of the transition region,  we expect that the transition region will be described by similarity solutions of the form
\begin{equation} \label{eq:simSols}
Q = (2\alpha)^{4/3}F_0^{1/3}z^{5/3}q(\eta), 
\quad
M = (2\alpha)^{2/3}F_0^{2/3}z^{4/3}m(\eta), 
\quad
F = F_0f(\eta),
\end{equation}
where the functions  $q$, $m$ and $f$ vary across the transition region, and where $F_0$ is the initial source buoyancy flux, $F_1$ is the new source buoyancy flux, $\alpha$ is the entrainment coefficient and $\eta$ is the nondimensional similarity variable defined by 
\begin{equation}
\eta = \frac{(2\alpha)^{1/2}}{F_0^{1/4}}\frac{z}{t^{3/4}}.
\end{equation}
This similarity variable, $\eta$, is analogous to that identified by \citet{woodhouseEtAl} but preserves a linear relationship between height and $\eta$.  The boundary conditions require that $q(0)=q_0(F_1/F_0)^{1/3}$, $m(0)=m_0(F_1/F_0)^{2/3}$, $f(0)=F_1/F_0$ while in the far-field, as $\eta \rightarrow \infty $, $q \rightarrow q_0$, $m \rightarrow m_0$ and $f\rightarrow 1$, where $q_0$ and $m_0$ are the nondimensional factors defined in \eqref{eq:q0m0} associated with steady plumes.

The similarity solutions are therefore governed by the coupled, non-linear, dimensionless set of ordinary differential equations
\begin{subequations} \label{eq:simEqs}
\begin{equation} \label{eq:massSim}
-\frac{3\eta^{7/3}}{4}\dv{}{\eta}\left(\frac{q^2}{m}\right) + \frac{1}{\eta^{2/3}}\dv{}{\eta}\left(\eta^{5/3}q\right) = m^{1/2} + \frac{\beta^*}{\eta^{2/3}}
\dv{}{\eta}\left[\eta^{2/3}m^{1/2}\dv{}{\eta}\left(\frac{\eta^2q^2}{m}\right)\right],
\end{equation}
\begin{equation} \label{eq:momSim}
-\frac{3\eta^{7/3}}{4}\dv{q}{\eta} + \frac{S_m}{\eta^{1/3}}\dv{}{\eta}\left(\eta^{4/3}m\right) = \frac{qf}{m} + \frac{\beta^*}{\eta^{1/3}}
\dv{}{\eta}\left[\eta^{2/3}m^{1/2}\dv{}{\eta}\left(\eta^{5/3}q\right)\right],
\end{equation}
\begin{equation} \label{eq:buoySim}
-\frac{3\eta^{4/3}}{4}\dv{}{\eta}\left(\frac{qf}{m}\right) + S_f\dv{f}{\eta} = \beta^*\dv{}{\eta}\left[\eta^{2/3}m^{1/2}\dv{}{\eta}\left(\eta^{1/3}\frac{qf}{m}\right)\right].
\end{equation}
\end{subequations}

We can solve these equations numerically, and determine solutions for $q, m$ and $f$ as functions of $\eta$ for given values of $\alpha$, $\beta$ and $F_1/F_0$.  The boundary value problem defined by \eqref{eq:simEqs} and the stated boundary conditions on the domain $[0, \infty)$ was solved by formulating the problem in terms of Chebyshev polynomials \citep[see][for details]{driscollEtAl}.  By comparison of the model prediction of the transition region with the data presented by \citep{scd08, sac09} for plumes in which there is (i) a decrease in the buoyancy flux and (ii) an increase in the buoyancy flux, we can then obtain an estimate for $\beta$. 

In comparing the model with the experimental observations we follow the evolution of the concentration of fluorscein  dye integrated across a horizontal plane through the centre of the plume.  The evolution of the concentration of the dye is analogous to the evolution of the buoyancy, as may be seen by noting that if a passive tracer has flux $C = C_0 c(\eta)$ (cf.~\ref{eq:simSols}$c$) then  $c$ is governed by an equation identical to (\ref{eq:simEqs}$c$) with $f$ replaced by $c$. 


\section{Comparison of solutions with experimental data}


\subsection{A decrease in buoyancy flux}


In \citet{scd08} a series of experiments describing the effect of a rapid reduction in the source buoyancy flux of a plume were presented.  A strong initial turbulent plume was established via two supplies of buoyant fluid to a single source.  At $t = 0$\,s one of the supplies of buoyant fluid was switched off, reducing the buoyancy flux at the source.  This reduction in buoyancy flux was due to a reduction in the source velocity.  The source diameter, density difference (reduced gravity) and tracer concentration (fluoroscein) were unchanged.  Data was presented from 100 experiments in which observations of the horizontally averaged concentration of dye as a function of height were made. By averaging the data from the nominally identical experiments, an ensemble average of the concentration as a function of height was produced at a series of fixed times after the change in the buoyancy flux.  The flux of tracer concentration at the source may be written as $C_0 = c_0 b_0^2 U_0$ and the source buoyancy flux is given by $G'_0 b_0^2 U_0$, where $c_0$ is the source concentration of tracer (not to be confused with the similarity solution for concentration {\it flux}, $c$), $G'_0$ is the reduced gravity at the source, $b_0$ is the source radius and $U_0$ is the source velocity.  Hence, reducing the buoyancy flux at the source by reducing the source velocity, but maintaining both the reduced gravity (density difference or buoyancy equivalently) and the source diameter, leads to a reduction in the source concentration flux, $C_0$, of exactly the same proportion.  For the experiments described in \citet{scd08}, the source buoyancy flux and hence concentration flux was reduced by a factor of 4.90.  The concentration of tracer at a given height, $z$, in the plume, $c/q$ is proportional to $C_0F_0^{-1/3}z^{-5/3}$.  Hence when $F_0$ is reduced by a factor of 4.90, the concentration is correspondingly reduced by a factor of $4.90^{2/3}\approx2.88$.

By rescaling the vertical coordinate with $t^{3/4}$, the concentration $c/q$ can be presented as a function of $\eta$, and if the transition is self-similar, the profiles should collapse to a universal curve. In figure \ref{fig:data}a, we present the experimental data, rescaled in this manner, for times $t= 0.83$\,s -- $16.29$\,s after the change in the buoyancy flux.  The grey shaded zone represents one standard deviation of the concentration relative to the mean as a function of $\eta$. It is seen that the concentration profile does collapse to a fixed profile, with the standard deviation being on average $1\%$, and no more that $4\%$, of the signal.  We consider the experimental data in a fixed interrogation window where the response of the light sheet to tracer concentration is linear and we observe the anticipated reduction in tracer concentration after the transient has passed.  The interrogation window had a vertical extent of 5\,cm and was centred on the axis of the plume at a height of 40\,cm above the plume source.  The design of the experiments means that there is a necessary change in the non-zero virtual origin \citep[see][]{huntKaye} between the initially forced jet-like and subsequently distributed, or `lazy', plume source conditions.  Here we assume that these effects are negligible within the interrogation window, far from the source, for comparison with the similarity solutions that satisfy \eqref{eq:simEqs}.

We have solved the new model equations as presented in \eqref{eq:sys} and now compare these solutions with the experimental data.   The nondimensional boundary condition for the source buoyancy flux that corresponds to the dimensional parameters in the experiments is $F_1/F_0 = 4.90^{-1} = 0.20$, and, following \citet{woodhouseEtAl}, values for the shape factor of both $S_m = 1.08$ and $S_m = 1.00$ were used together with $S_f = 1.00$ to assess the relative importance of the dispersion and the shear.  The value of $\beta^*$ was varied over the range 0 and $5/24$ and the root mean square (RMS) error between the model similarity solution and the experimental data calculated.  In figure \ref{fig:data}a, we illustrate the best fit model prediction for concentration, $c/q$, as a function of  $\eta$, and this is given by $\beta^*=0.14$.  The model prediction for $S_m = 1.08$ is shown solid, and the model prediction  for $S_m = 1.00$ is shown dashed, though it is difficult to distinguish the solutions at this scale. The RMS error between the model prediction and the experimental data as $\beta^*$ is varied is shown in figure \ref{fig:data}c for $S_m = 1.08$ (blue solid line) and $S_m = 1.00$ (blue dashed line).  In both cases the error is minimized for $\beta^* \approx 0.14$.  We observed that as $\beta$ decreased to very small values, the model predicts a sharp transition region  as the model prediction attempts to adjust from one buoyancy flux to another, somewhat reminiscent of the shock-like solutions presented by \citet{woodhouseEtAl}. However, for larger values of $\beta$ this transition region is predicted to spread  out vertically, and the model prediction  provides a closer fit to the experimental data. 

It is of relevance to note that \citet{woodhouseEtAl} proposed that the transition region spreads in time in a self-similar fashion owing to the different advection speeds of the momentum and buoyancy fluxes. These different speeds are associated with the different shape factors for the velocity and buoyancy profile in the plume.  The difference in the shape factor does indeed lead to a gradual spreading of the transition region, but, in the absence of the dispersion modelled herein, the model also  leads to the prediction of discontinuities in the concentration profile that were not observed in the experiments. Owing to the nonlinearity of the governing equations, the transition region spans a finite range of values of $\eta$ and within this region the plume adjusts smoothly from the original to the new steady buoyancy flux.  In figure \ref{fig:data}d we illustrate the minimum and maximum values of $\eta$, referred to as $\eta_1$ and $\eta_2$ respectively, across the transition zone as a function of $\beta^*$. We present solutions for the case in which the advection speed of momentum and buoyancy are the same (dashed), corresponding to the case $S_m= S_f =1$, and also for the case in which $S_m=1.08$, $S_f=1$ (solid), values consistent with experimental data for plumes with a constant buoyancy flux \citep[cf.][]{woodhouseEtAl, turner79}. It is seen that when $\beta^*$, and hence $\beta$, is very small, the two models lead to somewhat different ranges for $\eta$ across the transition zone. However,  as $\beta$ approaches the range of values which provide the best fit with the experimental data of \citet{scd08} ($\beta^*=0.14$), there is less than 1.2$\%$ difference in range $\eta_2 - \eta_1$ across the transition region between the model predictions using the different values for the shape factors. This  suggests that the effect of the different advection speeds of momentum and buoyancy  in causing the transition zone to spread out  is somewhat secondary to the effect of the dispersion, as parameterised by $\beta$. 


\subsection{An increase in buoyancy flux}


In \citet{sac09} the results of a series of numerical experiments in which the source buoyancy flux of an established plume was suddenly increased  were reported.  These data provide a second, and independent test of the similarity model \eqref{eq:simEqs}.  Again, the data describing the vertical variation of the horizontal average of the concentration of passive tracer collapses to a universal curve when we rescale height to the similarity variable  $\eta = z/t^{3/4}$.  In these numerical experiments the buoyancy flux was increased by simultaneously changing both the  reduced gravity and the velocity of the source fluid.  This was performed in such a way that the buoyancy flux at the source increased by a factor of 20, but the ratio of reduced gravity of the source fluid to the square of the velocity of the source fluid  was constant.  This implies that the source velocity was increased by a factor of $20^{1/3}$ and the source reduced gravity was increased by a factor of $20^{2/3}$.  This choice leads to a constant value $\varGamma_0 = 0.26$ so that the location of the virtual origin of the plume is a constant \citep[cf.][]{huntKaye}.  A consequence of this choice of boundary conditions is that for a steady plume, the tracer concentration at a height $z$ scales as $U_0^{2/3}G'_0\,\!^{-1/3}z^{-5/3}$. As a result, the expected  concentration at each height $z$ returns to its initial value after the transient has passed through.

\begin{figure}
\begin{center}
\includegraphics{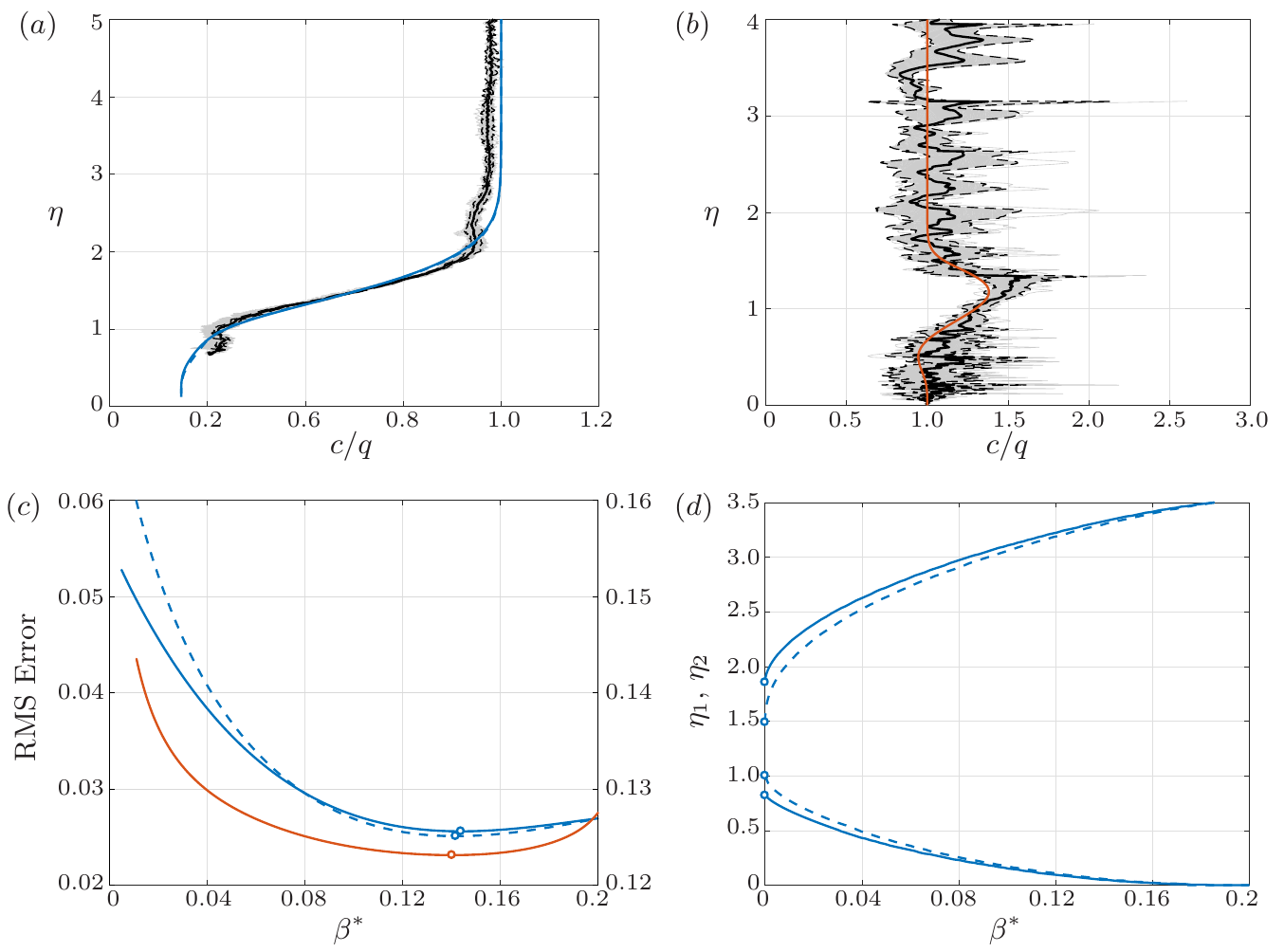}
\caption{\label{fig:data}(a) The experimental plume data from \citet{scd08} vs.~solution of \eqref{eq:simEqs}.  The individual rescaled experimental concentration contours from each time step in the ensemble are in grey.  A standard deviation either side of the mean is shown (black dashed) and the mean is the solid black line.  The best theoretical fit, corresponding to $\beta^* = 0.14$ is shown in solid blue for $S_m = 1.08$ and dashed blue for $S_m = 1.00$.  (b) The numerical plume data from \citet{sac09} vs.~solution of \eqref{eq:simEqs}.  The individual rescaled numerical concentration contours are in grey.  A standard deviation either side of the mean is shown (black dashed) and the mean is the solid black line.  The best theoretical fit, corresponding to $\beta^* = 0.14$ is shown in solid red for $S_m = 1.08$.  (c) The RMS Error for varying $\beta^*$ between the theoretical model and the experimental data (buoyancy turn-down, left-hand vertical axis) for $S_m = 1.08$ (solid blue), $S_m = 1.00$ (dashed blue) and between the theoretical model and the numerical data (buoyancy turn-up, right-hand vertical axis) in red.  (d)  The range over which the plume radius varies by a factor of $1 \pm 10^{-5}$ of its steady value $S_m = 1.08$ (blue solid) and $S_m = 1$ (blue dashed).  The lower two curves are $\eta_1$ and the upper two curves are $\eta_2$.  The solid data points, corresponding to $\beta^*=0$ are given by the similarity solutions of \citet{woodhouseEtAl} (blue solid) and the solutions of \citet{scase06} (blue dashed). As $\beta^*$ approaches the experimentally determined value of approximately $0.14$ the effect of $S_m$ being greater than 1 becomes negligible.}
\end{center}
\end{figure}

Figure \ref{fig:data}b illustrates the collapse of the experimental data to a universal curve, again with the grey zone representing the standard deviation of the concentration relative to the ensemble average at each height, as derived from the numerical simulations. The red curve is the best-fit prediction of the model \eqref{eq:simEqs} for the concentration as a function of $\eta$, optimized over $\beta^*\in[0, 5/24]$. Again, with small values of $\beta^*$ a shock like feature develops in the model predictions, but as $\beta^*$ increases to values of order $0.14$, the comparison with the data improves, as shown in figure \ref{fig:data}c (red) where we illustrate the RMS error between the numerical data and the solutions of \eqref{eq:simEqs}.

There is a significant overlap in the best fit range of values of $\beta^*$ for the cases of increasing and decreasing buoyancy flux, with the optimal value for both sets of data being $0.14$.  For $\alpha$ in the range $0.08$--$0.10$ this value of $\beta^*$ corresponds to values of $\beta$ in the range 0.70--0.88. 

\section{\label{sec:conc}Discussion and Conclusions}


The quantitative description of the  evolution of a turbulent buoyant plume with a variable source flux has attracted a considerable amount of attention, motivated by the modelling and experimental data presented by \citet{scase06, scld07, scd08, sac09} in which the transition in the structure of a plume as the buoyancy flux is either increased or decreased  was described. New analysis of this data, as presented herein, suggests that the transition in the plume is relatively smooth, and involves a region whose vertical extent grows in time at a rate proportional to $F_0^{1/4} t^{3/4}{\cal F}(F_0/F_1)$, where ${\cal F}(F_0/F_1)$ is a function of the initial to final buoyancy fluxes. Guided by this data, we propose a new model for the time dependent evolution of a turbulent buoyant plume, in which the mass, momentum and buoyancy are assumed to disperse along axis, with a turbulent dispersion coefficient $\beta U b$. We show the new model is consistent with the classical solutions for a steady turbulent buoyant plume, and that it admits self-similar solutions for the evolution of the transition zone in a plume if the flux is adjusted from one value to another. By comparison with the experimental data of \citet{scd08, sac09} we predict that $\beta$ has a value in the range 0.70--0.88. These solutions represent the long-term asymptotic flow in the case that the flux steadily adjusts from one value to another over a finite time.  We plan to explore the case in which the flow continually increases or decreases with time, in which case new experiments are required to describe the dynamics of both the associated starting plume \citep[cf.][]{turner62} and the time-evolving flow at each point below the starting plume \citep[cf.][]{deli79}.

\end{document}